Recurrent Stochastic Fluctuations with Financial Speculation[1]

Tomohiro Hirano[2]

Forthcoming at Festschrift in Honor of Joseph Stiglitz

**Section 1. Introduction**

Throughout history, many countries have repeatedly experienced large swings in asset prices, which are usually accompanied by large fluctuations in macroeconomic activity. One of the characteristics of the period before major economic fluctuations is the emergence of new financial products; the situation prior to the 2008 financial crisis is a prominent example of this. During that period, a variety of structured bonds, including securitized products, appeared. Because of the high returns on such financial products, many economic agents were involved in them for speculative purposes, even if they were riskier, producing macro-scale effects.

With this motivation, we present a simple macroeconomic model with financial speculation. Our model illustrates two points. First, stochastic fluctuations in asset prices and macroeconomic activity are driven by the repeated appearance and disappearance of risky financial assets, rather than expansions and contractions in credit availability. Second, in an economy with sufficient borrowing and lending, the appearance of risky financial assets leads to decreased productive capital, while in an economy with severely limited borrowing and lending, it leads to increased productive capital.

**Related literature**

The modern macro-finance literature including seminal papers by Stiglitz and Weiss (1981), Bernanke (1983), and Greenwald, Stiglitz, and Weiss (1984) emphasizes the role of credit availability for business fluctuations. Since these fundamentally important work, a large number of subsequent papers have been produced to the date, including Stiglitz and Weiss (1986, 1992), Woodford (1988), Bernanke and Gertler (1989), Greenwald and Stiglitz (1993), Kiyotaki and Moore (1997), Bernanke, Gertler, and Gilchrist (1999), Stiglitz and Greenwald (2003), and Brunnermeier and Sannikov (2014), among others.[3] In this literature, expansions and contractions in credit availability causes macroeconomic fluctuations. By contrast, in our model, the recurrent appearance and disappearance of risky financial assets produces large-scale and stochastic macroeconomic fluctuations.

---

[1] I am honoured to write this chapter for Festschrift in Honor of Joseph Stiglitz. Here I would like to take this opportunity to express my gratitude to Joe. I have been lucky enough to have the opportunity to write papers together and through countless numerous discussions, I have learnt more than words can describe. I also thank two anonymous referees for useful comments and the discussions with participants at the conference in Milan on May 24-25, 2023.

[2] First version, November 1st 2023, This version, August 8th 2024. Royal Holloway, University of London, and Center for Macroeconomics at the London School of Economics, and Canon Institute for Global Studies. Contact address: tomohih@gmail.com

[3] See the 2022 Nobel Memorial Prize in Economics for a comprehensive literature review. https://www.nobelprize.org/uploads/2022/10/advanced-economicsciencesprize2022-2.pdf

In terms of financial speculation and macroeconomic instability, our paper is related to Hirano and Stiglitz (2022), who develop a model of macroeconomic fluctuations with rational exuberance and land speculation. In their model, the economy sets off on unsustainable land-boom paths temporarily. There is a tipping point when land speculation reaches an unsustainable level, and bullish expectations suddenly must turn bearish, leading to an endogenous crash. In the long run, the macro-economy endogenously wobbles without converging to a steady state. A critical difference from Hirano and Stiglitz (2022) is that in our model, the equilibrium dynamics is uniquely determined globally, while in theirs, there are a plethora of rational expectation equilibria trajectories driven by people's beliefs.

In addition, regarding financial speculation, our paper is related to the so-called "rational bubble" literature that studies bubbles as speculation backed by nothing.[4] A critical difference from the rational bubble literature is that in our model, new financial products appear and disappear repeatedly, which drives stochastic macroeconomic fluctuations, but asset prices reflect fundamentals.

Our paper is also related to Guzman and Stiglitz (2021) in that fluctuations in wealth create macroeconomic instability. There are two main differences. First, in our model, the appearance and disappearance of risky financial assets produce fluctuations in wealth, but risky financial assets are connected to real assets generating dividends, while in Guzman and Stiglitz (2021), wealth is divorced from them. Second, our model is a full macro model with investment and production, while they consider an endowment economy.

**Section 2. The Model**

**2.1 The basic framework**

The model we consider is based on Hirano and Yanagawa (2017).[5] Consider a discrete-time economy with one homogeneous good and a continuum of entrepreneurs and workers with a unit measure, respectively. A typical entrepreneur and a representative worker have the following expected discounted utility.

(1) $E_t[\sum_{s=0}^{\infty}(\beta'\gamma)^s \log(c_{t+s}^i)]$,

where $i$ is the index for each entrepreneur and worker, and $c_t^i$ is the consumption of him/her at date $t$. $\beta'$ is the subjective discount factor, and $E_t[\cdot]$ is the expected value of $\cdot$ conditional on information at date 0. $\gamma$ is a survival rate of each agent. Each entrepreneur faces death shocks at a probability $\gamma < 1$ in each period, in which case (s)he derives no utility. Upon death new entrepreneurs with $1 - \gamma$ measure are born, so that the total population of entrepreneurs is constant with a unit measure. We also assume that upon death government seizes all the assets of entrepreneurs who died, and each new entrepreneur is equally endowed with the average wealth level of entrepreneurs who died. On the other hand, we assume for simplicity that workers live infinitely, i.e., $\gamma = 1$. We define $\beta'\gamma \equiv \beta$.

---

[4] The pioneering work includes Samuelson (1958), Shell, Sidrauski, and Stiglitz (1969, section 3), Bewley (1980), Tirole (1985), Scheinkman and Weiss (1986), Kocherlakota (1992), and Santos and Woodford (1997) that study pure bubbles, i.e., assets with no dividends like fiat money or cryptocurrencies (see Hirano and Toda 2024a for the literature review on pure bubble models). A series of recent papers by Hirano and Toda (2024a section 5 and 6, 2024b, 2024c, 2024d) and Hirano, Jinnai, and Toda (2024e) develops a theory of rational bubbles attached to real assets such as stocks, housing, or land yielding positive dividends. They derive that economic implications are markedly different between pure bubbles and bubbles attached to real assets.
[5] In the present model, risky financial assets generate positive dividends, while Hirano and Yanagawa (2017) study assets with no dividends, i.e., pure bubbles.

Each worker is endowed with one unit of labor endowment in each period, which is supplied inelastically in labor markets, and earns wages. We assume that workers are hand-to-mouth at all dates. In the rest of our analysis, we focus on entrepreneurs' behaviour.

At each date, each entrepreneur (both the existing and newborn entrepreneurs) has investment projects to produce capital. The investment technologies are as follows.

(2) $k_{t+1}^i = \alpha_t^i z_t^i$,

where $\alpha_t^i$ is productivity of investments and $z_t^i$ is investment of goods. For analytical tractability, we assume that capital fully depreciates after production.

In each period, each entrepreneur meets high-productivity projects (H-projects) at a probability $p$, in which case $\alpha_t^i = \alpha^H$, while at a probability $1 - p$, each entrepreneur meets low-productivity projects (L-projects), in which case $\alpha_t^i = \alpha^L$. We assume $\alpha^H > \alpha^L$. The probability $p$ is exogenous, and independent across entrepreneurs and over time. Productivity of each entrepreneur, whether (s)he has H-projects or L-projects, is public information between date $t$ and $t + 1$. We call entrepreneurs with H-projects (L-projects) H-types (L-types).

We first consider an economy where no borrowing and lending are possible. So, entrepreneurs need to self-finance their investment projects. The purpose of this assumption is to illustrate the point that large-scale macroeconomic fluctuations are driven by the repeated appearance and disappearance of risky financial assets, rather than the expansion and contraction of credit availability. In a later section, we will introduce borrowing and lending.

In this economy, risky financial assets appear and disappear stochastically. Risky financial assets are backed by trees. Aggregate supply of trees, when they are alive, is assumed to be fixed, $X_t = X$. $P_t^x$ is the price of a unit of tree in terms of consumption goods. Each tree yields $h_t^x$ units of consumption goods as dividends in each period. At a probability $\pi$, $h_t^x = h_t > 0$, i.e., the tree yields positive dividends, in which case the tree is sold at a positive price, i.e., $P_t^x = P_t > 0$. $P_t$ is determined endogenously in equilibrium. On the other hand, at a probability $1 - \pi$, it produces nothing (no dividends), i.e., $h_t^x = 0$, in which case $P_t^x = 0$. $1 - \pi$ captures the probability that an aggregate shock will happen. Once $1 - \pi$ hits the economy, the tree dies and yields nothing forever. Hence, its price becomes zero forever.[6] After this aggregate shock, at a probability $\rho$, newborn entrepreneurs enter the economy with new types of trees including initial dividends.[7] $h_t$ and/or $\pi$ of the new trees may be different from those of the previous trees. The endowment of new trees can be interpreted as an appearance of new risky assets. Probabilities $\rho$ and $1 - \pi$ capture frequencies of the appearance and disappearance of risky financial products. We assume that only newborn entrepreneurs can create risky financial products backed by trees and no contingent claims to new financial assets that may pop up in the future can be made before their appearance.

---

[6] This assumption may be extreme but may have some relevancy. For instance, in the 2008 financial crisis, securitized assets like mortgage-backed securities were backed by the housing prices. The decline in the housing prices was an aggregate shock, by which those securitized assets backed by the housing prices produced almost no dividends, and the prices of those assets went nearly zero.

The entrepreneur's flow of funds constraint is given by

(3) $\quad c_t^i + z_t^i + P_t^x x_t^i = q_t \alpha_{t-1}^i z_{t-1}^i + (h_t^x + P_t^x) x_{t-1}^i \equiv e_t,$

where $x_t^i$ is the amount of trees purchased by a type $i$ entrepreneur at date $t$. The left-hand side of (3) is expenditure on consumption, investment, and the purchase of trees. The right-hand side is composed of two terms, i.e., the return from investment in the previous period and the return from holding trees (total dividends plus total returns from selling trees). We define the right-hand side as wealth of entrepreneur $i$.

We also impose the short sale constraint on trees.

(4) $\quad x_t^i \geq 0.$

## 2.2 Production

There are competitive firms which produce the final consumption goods using capital and labor. The production function of each firm $j$ is given by

(5) $\quad y_{tj} = k_{tj}^\sigma l_{tj}^{1-\sigma}.$

Factor prices satisfy the marginal product. Since aggregate labor supply of workers is one, we obtain

(6) $\quad q_t = \sigma K_t^{\sigma-1},$

(7) $\quad w_t = (1 - \sigma) K_t^\sigma,$

where $q_t$ and $w_t$ are the rental rate of capital and the wage rate, respectively. $K_t$ is aggregate capital stock at date $t$.

Aggregate production in this economy is given by

(8) $\quad Y_t = K_t^\sigma + h_t^x X.$

We assume that so long as trees are alive and yield dividends, productivity of trees grows with aggregate capital stock in the economy. This assumption ensures saddle point stability. We take a particular form.

(9) $\quad h_t = D K_t^\sigma.$

---

[7] Including initial dividends is a technical assumption. When we do not include them, the dynamic equations at date $t$ when new financial assets pop up and those at date $t + 1$ onwards are slightly different. Hence, the initial price adjustment will occur at date $t$ so that the economy will get on the saddle path at date $t + 1$ onwards. With the exception of this point, the analysis is the same.

When $\sigma = 1$, this model corresponds to the standard AK model. In the rest of our analysis, we mainly focus on the case with $\sigma < 1$.

### 2.3 Entrepreneurs' behaviour

Since we employ log-utility, each entrepreneur consumes a fraction $1 - \beta$ of his/her wealth, i.e., $c_t^i = (1 - \beta)e_t^i$. In equilibrium, H-types invest in their projects instead of buying risky financial assets because the expected rate of return on their own H-projects is strictly greater than that of risky financial assets, i.e., the short-sale constraint (4) binds. We will verify this in Appendix 1. That is,

(10) $\quad z_t^i = \beta e_t^i.$

On the other hand, L-types decide optimal portfolio allocations between holding risky financial assets and investing in their projects with safe returns. Solving optimal portfolio allocations yields the demand equation for risky financial assets.

(11) $\quad P_t x_t^i = \left(\frac{\pi R_t^x - q_{t+1}\alpha^L}{R_t^x - q_{t+1}\alpha^L}\right)\beta e_t^i,$

where $R_t^x \equiv \frac{h_{t+1}}{P_t} + \frac{P_{t+1}}{P_t}$ is the rate of return on holding risky financial assets as long as $\pi$ persists.

Since entrepreneurs are risk averse, L-types invest the remaining savings in their own projects with low return but safe assets. They do so for risk-hedge.

(12) $\quad z_t^i = \left(\frac{(1-\pi)R_t^x}{R_t^x - q_{t+1}\alpha^L}\right)\beta e_t^i.$

### 2.4 Definition of competitive equilibrium

Let us denote aggregate consumption and wealth of entrepreneurs at date $t$ as $\int_{i \in H_t \cup L_t} c_t^i di \equiv C_t$ and $\int_{i \in H_t \cup L_t} e_t^i di \equiv A_t$. Similarly, let $\int_{i \in H_t} z_t^i di \equiv Z_t^H$, $\int_{i \in L_t} z_t^i di \equiv Z_t^L$, and $\int_{i \in H_t \cup L_t} x_t^i di \equiv X_t$ be aggregate investments by H-types and L-types at date $t$, and aggregate demand for risky financial assets at date $t$. Then, the competitive equilibrium is defined as a set of prices $\{P_t, q_t, w_t\}_{t=0}^{\infty}$ and aggregate quantities $\{Z_t^H, Z_t^L, C_t, A_t, Y_t, K_{t+1}\}_{t=0}^{\infty}$, given an initial $K_0$, such that each entrepreneur chooses consumption, investments, and the amount of risky financial assets $\{c_t^i, z_t^i, x_t^i\}_{t=0}^{\infty}$ to maximize the expected discounted utility (1) subject to (2), (3), and (4), and all markets clear, including goods and risky assets markets. That is,

(13) $\quad C_t + Z_t^H + Z_t^L = Y_t.$

(14) $\quad X_t = X.$

### 2.5 Aggregation

We are now going to derive the aggregate dynamics when $\pi$ persists.

Aggregate wealth of entrepreneurs is given by

(15) $A_t = q_t K_t + (h_t + P_t)X = \sigma K_t^\sigma + (h_t + P_t)X$.

The aggregate wealth of entrepreneurs is composed of two terms, i.e., aggregate returns to capital $q_t K_t$ plus aggregate returns to holding risky financial assets $(h_t + P_t)X$.

By aggregating (11), we can derive the aggregate demand for risky financial assets.

(16) $P_t X = \left(\frac{\pi R_t^x - q_{t+1}\alpha^L}{R_t^x - q_{t+1}\alpha^L}\right)\beta(1-p)A_t$.

Rearranging the goods market clearing condition yields

(17) $Z_t^H + Z_t^L + P_t X = \beta A_t$.

Aggregate savings $\beta A_t$ finances aggregate capital investments, $Z_t^H + Z_t^L$, and the aggregate value of risky assets.

Then, we have the evolution of aggregate capital stock.

(18) $K_{t+1} = \alpha^H Z_t^H + \alpha^L Z_t^L = \alpha^H \beta p A_t + \alpha^L[\beta(1-p)A_t - P_t X]$.

From (18), we learn that there are two competing effects of a rise in $P_t X$ on aggregate capital. $(-P_t X)$ in the second term captures a crowding out effect. That is, without risky financial assets, L-types put all their savings into their L-projects but once risky financial assets pop up in the economy, their existence crowds their savings away from L-projects. On the other hand, if other things being constant, a rise in $P_t X$ increases both types of investments through raising aggregate wealth $A_t$, i.e., a wealth effect. This wealth effect crowds both investments in.

More precisely, using (15), (18) can be written as

(19) $K_{t+1} = \alpha^H \beta[\sigma K_t^\sigma + (h_t + P_t)X] + \alpha^L\{\beta(1-p)[\sigma K_t^\sigma + (h_t + P_t)X] - P_t X\}$.

By partially differentiating (19) with respect to $P_t X$, we have

(20) $\frac{\partial K_{t+1}}{\partial P_t X} = \alpha^H \beta + \alpha^L[\beta(1-p) - 1]$.

The first term shows that the rise in $P_t X$ (or the appearance of risky financial products) crowds H-projects in. On the other hand, the sign of the second term is negative. This means that although there are both crowding-in and crowding-out effects on L-projects, the overall effect is that the rise in $P_t X$ crowds L-projects out. Therefore, as an impact on the macro-economy, whether the appearance of risky financial assets is expansionary or contractionary depends on the size of the effect of increasing H-projects and decreasing L-projects. From (20), if $\alpha^H$ is large enough relative to $\alpha^L$, the crowding-in effect on H-projects is expected to dominate the crowding-out effect on L-projects.

## 2.6 Dynamics and stochastic steady states

To solve the model, we define $\phi_t$ as the size of financial speculation relative to aggregate savings. That is, $\phi_t \equiv \frac{P_t X}{\beta A_t}$.

Solving (16) for $R_t^x$ by using $\phi_t$ yields

(21) $R_t^x = \frac{h_{t+1} + P_{t+1}}{P_t} = \frac{R_{t+1} \alpha^L (1 - p - \phi_t)}{\pi(1-p) - \phi_t}.$

Using $\phi_t$ and (21), the growth rate of aggregate wealth can be written as

(22) $1 + a_t \equiv \frac{A_{t+1}}{A_t} = q_{t+1} \left\{ \left(1 + \left(\frac{\alpha^H - \alpha^L}{\alpha^L}\right) p\right) \beta \alpha^L - \beta \alpha^L \phi_t \right\} + \frac{q_{t+1} \alpha^L (1-p-\phi_t) \beta \phi_t}{\pi(1-p) - \phi_t}.$

The evolution of $\phi_t$ follows according to

(23) $\phi_{t+1} = \left(\frac{P_{t+1}/P_t}{1 + a_t}\right) \phi_t,$

i.e., it depends on the growth rates of asset prices and aggregate wealth.

(21) can be rewritten as

(24) $\frac{P_{t+1}}{P_t} = R_t^x - \frac{DK_{t+1}^\sigma}{P_t} = R_t^x - DX \frac{\beta A_t}{P_t X} \frac{A_{t+1}}{A_t} \frac{K_{t+1}^\sigma}{\beta A_{t+1}}.$

Since $A_t = \frac{(\sigma + DX) K_t^\sigma}{1 - \beta \phi_t}$, substituting this relation into (24) and then into (23) yields

(25) $\phi_{t+1} = \left(\frac{R_t^x}{1+a_t}\right) \phi_t \left(1 + \frac{DX}{\sigma}\right) - \frac{DX}{\beta \sigma}.$

Substituting (21) and (22) into (25) yields

(26) $\phi_{t+1} = \left( \frac{\frac{(1-p-\phi_t)}{\pi(1-p) - \phi_t}}{\left(1 + \left(\frac{\alpha^H - \alpha^L}{\alpha^L}\right) p\right) \beta \alpha^L - \beta \alpha^L \phi_t + \frac{(1-p-\phi_t) \beta \phi_t}{\pi(1-p) - \phi_t}} \right) \phi_t \left(1 + \frac{DX}{\sigma}\right) - \frac{DX}{\beta \sigma}.$

Also, by substituting $A_t = \frac{(\sigma + DX) K_t^\sigma}{1 - \beta \phi_t}$ into (18), we obtain the dynamic equation concerning aggregate capital stock.

(27) $K_{t+1} = \frac{\left(1 + \left(\frac{\alpha^H - \alpha^L}{\alpha^L}\right) p\right) \beta \alpha^L - \beta \alpha^L \phi_t}{1 - \beta \phi_t} (\sigma + DX) K_t^\sigma.$

Therefore, the dynamics of this economy can be characterized by (26) and (27).

We impose conditions on parameter values.

**Assumption 1.** $\pi \left(1 + \left(\frac{\alpha^H - \alpha^L}{\alpha^L}\right) p\right) \beta > 1 \leftrightarrow \frac{\alpha^H}{\alpha^L} > \frac{1 - \pi \beta (1-p)}{\pi p \beta}.$

This assumption ensures that even if $D = 0$, the stochastic steady state $\phi > 0$ exists.

From (26), we learn that $\phi_{t+1}$ is a convex function of $\phi_t$, with the negative intercept when $D > 0$. Hence, there exists a unique and positive value of $\phi$ where $\phi_t = \phi_{t+1} \equiv \phi$. Once $\phi_t$ remains at $\phi$, we learn from (27) that capital stock will converge to a stochastic steady state as long as $\pi$ persists. Note that in our model, stochastic steady state means a state that so long as $\pi$ or $1 - \rho$ persists, $K$ and other aggregate variables (consumption, investment, and wealth) will be constant over time.

The stochastic steady state with risky financial assets is given by

$$(28) \quad K = \left[\frac{\left(1+\left(\frac{\alpha^H-\alpha^L}{\alpha^L}\right)p\right)\beta\alpha^L - \beta\alpha^L\phi}{1-\beta\phi}(\sigma + DX)\right]^{\frac{1}{1-\sigma}}.$$

In the rest of our analysis, we focus on the case where $D$ is sufficiently small, i.e., $D \to 0$, so that at $\phi_t = \phi$, H-types invest in their own projects, instead of buying risky financial assets. We will show below that $\phi_t = \phi$ is the only rational expectations equilibrium trajectory.

First, let us analyze the economy without risky financial assets, which corresponds to the economy with $\phi_t = 0$ and $D = 0$. Then, (27) changes to

$$(29) \quad K_{t+1} = \left[\left(1 + \left(\frac{\alpha^H-\alpha^L}{\alpha^L}\right)p\right)\beta\alpha^L\right]\sigma K_t^\sigma.$$

Given an initial capital stock $K_0$, this economy will converge to a stochastic steady state without risky financial asset.

The stochastic steady state without risky financial assets is given by

$$(30) \quad K = \left[\left[\left(1 + \left(\frac{\alpha^H-\alpha^L}{\alpha^L}\right)p\right)\beta\alpha^L\right]\sigma\right]^{\frac{1}{1-\sigma}}.$$

To summarize the results above, we obtain the following Propositions.

**Proposition 1**: (Existence of stochastic steady states)

In this economy, there exist two stochastic steady states, one with risky financial assets, and the other without them. The values of capital stock at respective stochastic steady states are given by (28) and (30).

**Proposition 2**

(2-i) Dynamics of an economy without risky financial assets

Given an initial capital stock $K_0$, there exists a unique dynamic path converging to the stochastic steady state given by (30).

(2-ii) Dynamics of an economy with risky financial assets

Consider $D > 0$ sufficiently small, i.e., $D \to 0$. Under Assumption 1, given an initial capital stock $K_0$, the initial land prices $P_0$ will be instantly set to achieve $\phi$, without transitional dynamics. This is the only equilibrium trajectory consistent with rational expectations. Otherwise, land prices would explode or implode, which cannot be part of equilibrium trajectory. Once $\phi$ gets constant, the macro-economy will converge to the stochastic steady state given by (28) as long as $\pi$ persists.

## 2.7 Macroeconomic effects of risky financial assets

As long as $\pi$ persists, $\phi_t = \phi$. Then, (27) becomes

(31) $K_{t+1} = \frac{\left(1+\left(\frac{\alpha^H-\alpha^L}{\alpha^L}\right)p\right)\beta\alpha^L - \beta\alpha^L\phi}{1-\beta\phi}(\sigma + DX)K_t^\sigma.$

When we compare (29) to (31), for any value of $K_t$, we have

(32) $K_{t+1}$ in (31) $- K_{t+1}$ in (29) $= \frac{\left[\left(1+\left(\frac{\alpha^H-\alpha^L}{\alpha^L}\right)p\right)\beta - 1\right]\beta\alpha^L\phi}{1-\beta\phi}\sigma K_t^\sigma +$
$\frac{\left(1+\left(\frac{\alpha^H-\alpha^L}{\alpha^L}\right)p\right)\beta\alpha^L - \beta\alpha^L\phi}{1-\beta\phi}DXK_t^\sigma > 0,$

because $\left(1+\left(\frac{\alpha^H-\alpha^L}{\alpha^L}\right)p\right)\beta - 1 > 0$ and $\frac{\left(1+\left(\frac{\alpha^H-\alpha^L}{\alpha^L}\right)p\right)\beta\alpha^L - \beta\alpha^L\phi}{1-\beta\phi} > 0$ under Assumption 1 and if $D \to 0$ (see also Appendix 1). That is, the appearance of risky financial assets generates expansionary effects on capital stock. We summarize this result in the following Proposition.

**Proposition 3** (Macroeconomic impact of an appearance of risky financial assets)

For $D > 0$ sufficiently small, i.e., $D \to 0$, under Assumption 1, the crowding-in effect on H-projects dominates the crowding-out effect on L-projects. Therefore, the appearance of risky financial assets generates expansionary effects on aggregate capital stock, aggregate output, aggregate wealth, aggregate consumption, and wages.

Figure 1 below illustrates this situation. Points A and B are stochastic steady states of the economy with and without risky financial products, respectively. Intuitively, without risky financial products, L-types put all their savings in their own projects with low returns. Once risky financial products pop up, L-types purchase them and sell them when they have H-projects in the future. Since risky financial assets yield greater returns than L-projects, this increases the returns on savings for L-types, leading to increased aggregate wealth and financing more H-projects, while L-projects are crowded out. This reallocation of resources towards H-projects has an expansionary effect on the macro-economy.

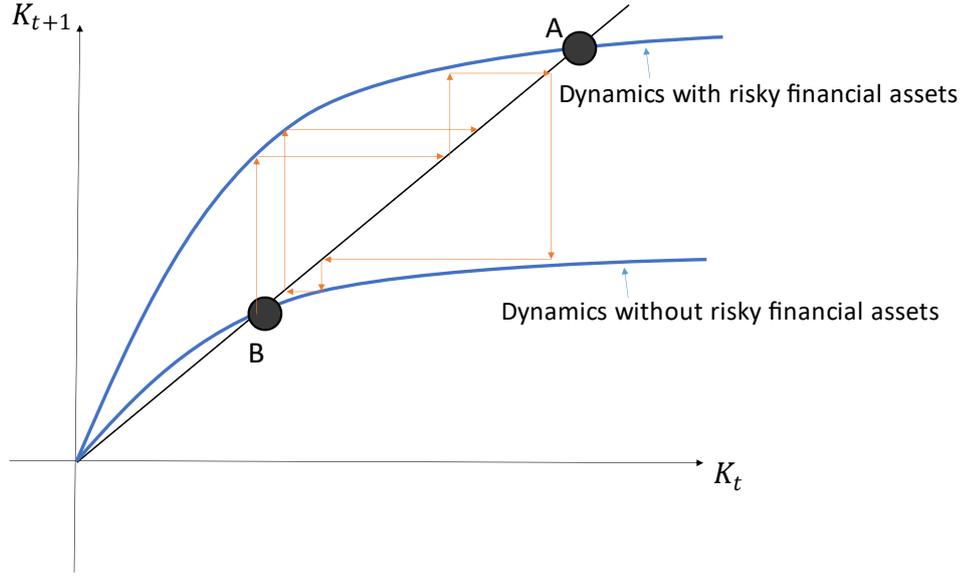

Figure 1: Dynamics with and without risky financial assets

Moreover, Proposition 3 directly leads to the following Proposition.

**Proposition 4** (Recurrent stochastic fluctuations with financial speculation)

Consider the economy described above. The equilibrium dynamics is uniquely determined globally. Depending on probabilities $\rho$ and $\pi$, the macro-economy fluctuates between points A and B in the long run recurrently and stochastically. That is, when risky financial assets pop up, their existence produces economic booms in macroeconomic variables, while their disappearance leads to economic contractions. This process repeats itself stochastically.

**2.8 The case with $\sigma = 1$**

When $\sigma = 1$, there exist two stochastic balanced growth paths.

The growth rate with risky financial assets is given by

$$\frac{K_{t+1}}{K_t} = \frac{\left(1+\left(\frac{\alpha^H-\alpha^L}{\alpha^L}\right)p\right)\beta\alpha^L - \beta\alpha^L\phi}{1-\beta\phi}(1+DX).$$

The growth rate without them is given by

$$\frac{K_{t+1}}{K_t} = \left[\left(1+\left(\frac{\alpha^H-\alpha^L}{\alpha^L}\right)p\right)\beta\alpha^L\right].$$

From the discussion in the previous section, it is straightforward that the growth rate with risky financial assets is higher than that without them. Therefore, the macro-economy experiences recurrent ups and downs in economic growth stochastically, together with the appearance and disappearance of risky financial assets.

### Section 3. Introducing borrowing and lending

### 3.1 An extended framework with borrowing and lending

We now introduce borrowing and lending to the basic model.

The entrepreneur's flow of funds constraint changes to

(33) $c_t^i + z_t^i + P_t^x x_t^i = q_t \alpha_{t-1}^i z_{t-1}^i + (h_t^x + P_t^x) x_{t-1}^i - (1 + r_t) b_t^i \equiv e_t,$

where $b_t^i > 0 \ (< 0)$ is borrowing (lending), and $1 + r_t$ is the gross interest rate between date $t$ and $t + 1$.

We consider the following borrowing constraint.

(34) $(1 + r_t) b_t^i \leq \theta q_{t+1} \alpha_t^i z_t^i,$

i.e., only a fraction $\theta \in [0,1]$ of the return from investment can be used as collateral. $\theta$ captures the pledgeability value in borrowing. $\theta = 0$ corresponds to the basic model.

When the borrowing constraint (34) binds, the investment function of each entrepreneur can be written as

(35) $z_t^i = \dfrac{\beta e_t^i}{1 - \dfrac{\theta q_{t+1} \alpha_t^i}{1+r_t}},$

i.e., it equals leverage $1/\left(1 - \dfrac{\theta q_{t+1} \alpha_t^i}{1+r_t}\right)$ times savings $\beta e_t^i$.

On the other hand, if the borrowing constraint does not bind, entrepreneurs decide optimal portfolio allocations between $z_t^i$, $x_t^i$, and $b_t^i$. Solving the optimal allocations yields the demand equation for risky financial assets.

(36) $P_t x_t^i = \left(\dfrac{\pi R_t^x - (1+r_t)}{R_t^x - (1+r_t)}\right) \beta e_t^i.$

The remaining savings is split between $z_t^i$ and $b_t^i$. That is,

(37) $z_t^i + (-b_t^i) = \left(\dfrac{(1-\pi) R_t^x}{R_t^x - (1+r_t)}\right) \beta e_t^i.$

Since $z_t^i$ and $b_t^i$ are both safe assets, $z_t^i = 0$ if $1 + r_t > q_{t+1} \alpha_t^i$, while $z_t^i \geq 0$ if $1 + r_t = q_{t+1} \alpha_t^i$. Note that $1 + r_t < q_{t+1} \alpha^L$ cannot be equilibrium because all entrepreneurs would want to borrow, and no one would become lenders.

In the rest of our analysis, we focus on the following case to illustrate the point that unlike the case of $\theta = 0$, when the pledgeability value is high enough, the appearance of risky financial assets ends up crowding H-projects out, rather than crowding them in, thereby leading to decreased productive capital.

**Assumption 2.** $\theta \geq \frac{\alpha^L}{\alpha^H}(1-p)$

Under Assumption 2, since the pledgeability value is high enough, H-types can borrow sufficiently that the equilibrium interest rate will become greater than the rate of return on L-projects. As a result, L-types have no incentives to invest in their projects. The purpose of this assumption is to show the point that in an economy with sufficient borrowing and lending, the appearance of risky financial assets leads to decreased productive capital by crowding it out.

### 3.2 Economy without risky financial assets

Let us first consider the economy without risky financial assets.

Since only H-types invest, the evolution of aggregate capital follows according to

(38) $\quad K_{t+1} = \alpha^H \beta A_t = \alpha^H \beta \sigma K_t^\sigma.$

Given an initial capital stock $K_0$, the macro-economy converges to a stochastic steady state given by

(39) $\quad K = (\alpha^H \beta \sigma)^{\frac{1}{1-\sigma}}.$

**Proposition 5** (Existence of a stochastic steady state in an economy with no risky financial assets but high $\theta$)

Under Assumption 2, there exists a unique stochastic steady state given by (39). Given an initial capital stock $K_0$, there is a unique dynamic path converging to the stochastic steady state.

### 3.3 Economy with risky financial assets

When the borrowing constraint of H-types binds (this will be the case in equilibrium), their aggregate investment function is

(40) $\quad Z_t^H = \frac{\beta A_t}{1 - \frac{\theta q_{t+1} \alpha^H}{1+r_t}} = \frac{\beta [\sigma K_t^\sigma + (h_t + P_t)X]}{1 - \frac{\theta q_{t+1} \alpha^H}{1+r_t}}.$

A rise in $P_t X$ or an appearance of risky financial assets produces the wealth effect, thereby financing more H-projects, i.e., the crowding-in effect is generated. With leverage, this effect is amplified.

The equilibrium interest rate is determined so that aggregate borrowing and lending are equalized. That is,

(41) $\frac{\beta p A_t}{1-\frac{\theta q_{t+1}\alpha^H}{1+r_t}} + P_t X = \beta A_t.$

Aggregate savings $\beta A_t$ finances H-projects, which is the first term on the left-hand side, and the aggregate value of risky financial assets, which is the second term. Other things being constant, a rise in $P_t X$ crowds savings away from H-projects, in other words, the equilibrium interest rate goes up, making financing H-projects difficult and generating the crowding-out effect. The overall effect with the rise in $P_t X$ on H-projects depends on whether which one of the crowding-out or the crowding-in effects dominates.

Solving (41) for $1 + r_t$ yields

(42) $1 + r_t = \frac{\theta q_{t+1}\alpha^H(1-\phi_t)}{1-p-\phi_t}.$

Other things being constant, a rise in $\phi_t$ leads to higher interest rates because more funds flow to financial speculation, which tightens the borrowing and lending market.

Aggregating (36) and then solving for $R_t^x$ using (42), we obtain

(43) $R_t^x = \frac{(1+r_t)(1-p-\phi_t)}{\pi(1-p)-\phi_t} = \frac{\theta q_{t+1}\alpha^H(1-\phi_t)}{\pi(1-p)-\phi_t}.$

The growth rate of aggregate wealth $A_{t+1} = q_{t+1}\alpha^H Z_t^H + R_t^x P_t X$ can be written as

(44) $1 + a_t \equiv \frac{A_{t+1}}{A_t} = q_{t+1}\alpha^H(1-\phi_t)\frac{\pi(1-p)-(1-\theta)\phi_t}{\pi(1-p)-\phi_t}.$

By substituting (43) and (44) into (25), we obtain the evolution of $\phi_t$.

(45) $\phi_{t+1} = \frac{\theta}{\beta}\frac{\phi_t}{\pi(1-p)-(1-\theta)\phi_t}\left(1+\frac{DX}{\sigma}\right) - \frac{DX}{\beta\sigma}.$

The evolution of aggregate capital follows according to

(46) $K_{t+1} = \alpha^H Z_t^H = \alpha^H(1-\phi_t)\beta A_t = \frac{\alpha^H\beta(1-\phi_t)}{1-\beta\phi_t}(\sigma + DX)K_t^\sigma.$

Therefore, the dynamics of this economy can be characterized by (45) and (46).

We impose the following conditions on parameter values.

**Assumption 3.** $\theta < \pi\beta(1-p)$

This Assumption 3 ensures that even if $D = 0$, the stochastic steady state $\phi > 0$ exists. Assumptions 2 and 3 mean that we consider the region of $\frac{\alpha^L}{\alpha^H}(1-p) \leq \theta < \pi\beta(1-p)$. When $\pi$ and $\beta$ are close to one and $p$ is sufficiently small, the upper bound approaches nearly one. Moreover, if $\alpha^H$ is large enough, this region is not empty.

As in the basic model, in (45), $\phi_{t+1}$ is a convex function of $\phi_t$, with the negative intercept when $D > 0$. Hence, there exists a unique and stochastic steady state where $\phi_t = \phi_{t+1} \equiv \phi$ and $K$ becomes constant as long as $\pi$ persists. Aggregate capital at the stochastic steady state is given by

(47) $K = \left(\frac{\alpha^H(1-\phi)}{1-\beta\phi}(\sigma + DX)\right)^{\frac{1}{1-\sigma}}.$

**Proposition 6** (Existence of a stochastic steady state in an economy with risky financial assets and high $\theta$ and the economy's dynamics)

Consider $D > 0$ sufficiently small, i.e., $D \to 0$. Under Assumption 2 and 3, i.e., $\frac{\alpha^L}{\alpha^H}(1-p) \leq \theta < \pi\beta(1-p)$, given an initial capital stock $K_0$, the initial land prices $P_0$ will be instantly set to achieve $\phi$, without transitional dynamics. This is the only equilibrium trajectory consistent with rational expectations. Once $\phi$ becomes constant, the macro-economy will converge to the stochastic steady state given by (47) as long as $\pi$ persists.

Note that in $\frac{\alpha^L}{\alpha^H}(1-p) \leq \theta < \pi\beta(1-p)$, if $D \to 0$, the borrowing constraint for H-types will bind in the equilibrium trajectory and H-types do not buy risky financial assets. We verify this in Appendix 2.

When we compare (38) with (46) (in equilibrium $\phi_t = \phi$), for $D > 0$ sufficiently small, aggregate capital with risky financial assets is strictly lower than that without them. In other words, the crowding-out effect dominates the crowding-in effect. We summarize this result in the following Proposition.

**Proposition 7** (Macroeconomic effects of an appearance of risky financial assets in an economy with high $\theta$)

For $D > 0$ sufficiently small, i.e., $D \to 0$, when Assumptions 2 and 3 hold, i.e., $\frac{\alpha^L}{\alpha^H}(1-p) \leq \theta < \pi\beta(1-p)$, the crowding-out effect on H-projects dominates the crowding-in effect on them. Therefore, the appearance of risky financial assets leads to decreased productive capital.

This Proposition 8 implies that when we think about Figure 1, in the region of $\frac{\alpha^L}{\alpha^H}(1-p) \leq \theta < \pi\beta(1-p)$, the dynamics of this economy with and without risky financial assets is opposite to the dynamics of the economy with $\theta = 0$. This leads to the following Proposition.

**Proposition 8** (Recurrent stochastic fluctuations with financial speculation in an economy with high $\theta$)

Consider the economy described above, where $\frac{\alpha^L}{\alpha^H}(1-p) \leq \theta < \pi\beta(1-p)$. The equilibrium dynamics is uniquely determined globally. When risky financial assets pop up at a probability ρ, their existence crowds H-projects out, leading to decreased productive capital, while with their disappearance, more funds flow into productive capital. This process repeats itself stochastically.

### 3.3 The case with $\sigma = 1$

As in the basic model, there are two stochastic balanced growth paths.

The stochastic steady-state growth with risky financial assets is given by

$$\frac{K_{t+1}}{K_t} = \frac{\alpha^H \beta (1-\phi)}{1-\beta\phi}(1 + DX).$$

The stochastic steady-state growth without them is given by

$$\frac{K_{t+1}}{K_t} = \alpha^H \beta.$$

From the discussion in the previous section, it is straightforward that when $D \to 0$, the appearance of risky financial assets reduces economic growth so long as they exist.

Our analyses of when $\theta = 0$ and when $\theta$ is high enough, respectively, imply that macroeconomic effects of the appearance of risky financial assets depends on the value of $\theta$. By continuity, in economies where $\theta$ is low enough, the appearance of risky financial assets leads to increased productive capital, while in economies where $\theta$ is high enough, it leads to decreased productive capital.

Here we like to add differences from Hirano and Stiglitz (2024), who develops a two-period OLG model with credit frictions, endogenous growth, and two sectors. In their model, different assets (capital investments and land) have different collateral values and entrepreneurs are engaged in both capital investment and land speculation both using borrowing. Under such circumstances, they show that an increase in the land collateral value or low interest policies may end up crowding productive capital out, rather than crowding them in, reducing long-run economic growth. In our model, only returns from capital investment can be used as collateral, and in equilibrium, H-types do not participate in land speculation and only invest in capital. Still, when the collateral value is high enough, the existence of risky financial assets crowds productive capital out. Although the approaches are different, they are both complementary in understanding crowding-out effects of financial speculation.

**Appendices**

**Appendix 1**

We will verify $q_{t+1}\alpha^H > \pi R_t^x$ in equilibrium, i.e., the short-sale constraint (4) will bind in equilibrium.

As $D \to 0$,

$$(48) \quad \phi = \frac{\pi\left(1+\left(\frac{\alpha^H-\alpha^L}{\alpha^L}\right)p\right)\beta-1}{\left(1+\left(\frac{\alpha^H-\alpha^L}{\alpha^L}\right)p\right)\beta-1-\beta(1-p)(1-\pi)}.$$

By substituting (48) into (31), we have

$$(49) \quad K_{t+1} = \frac{\left(1+\left(\frac{\alpha^H-\alpha^L}{\alpha^L}\right)p\right)\beta\alpha^L-\beta\alpha^L(1-p)}{1-\pi\beta(1-p)} \sigma K_t^\sigma,$$

where the numerator is positive under Assumption 1.

Then, $q_{t+1}\alpha^H > \pi R_t^x$ can be written as

$$(50) \quad \alpha^H > \frac{\pi K_{t+1}}{\sigma K_t^\sigma} = \frac{\pi\left(1+\left(\frac{\alpha^H-\alpha^L}{\alpha^L}\right)p\right)\beta\alpha^L-\pi\beta\alpha^L(1-p)}{1-\pi\beta(1-p)},$$

which can be further written as

$$(51) \quad \alpha^H[1-\pi\beta(1-p)] > \alpha^H \pi p \beta,$$

which is true.

**Appendix 2**

The proofs here are basically the same as the ones in the Online Appendix 9.17 of Hirano and Yanagawa (2017).

First, we prove that the borrowing constraint for H-types binds in equilibrium, that is, $q_{t+1}\alpha^H > 1+r_t$.

Using (42), we obtain

$$(52) \quad q_{t+1}\alpha^H > 1+r_t = \frac{\theta q_{t+1}\alpha^H(1-\phi)}{1-p-\phi}.$$

As $D \to 0$, we obtain

$$(53) \quad \phi = \frac{\pi\beta(1-p)-\theta}{\beta(1-\theta)}.$$

By substituting (53) into (52) and rearranging it, (52) becomes

$$(54) \quad \beta(1-p)(1-\pi) + \theta(1-\beta) > 0,$$

which is true.

Next, we prove that H-types do not buy risky financial assets in equilibrium, i.e., the short-sale constraint (4) will bind in equilibrium.

If the leveraged rate of return from H-projects is strictly greater than the expected rate of return from risky financial assets, that is,

$$(55) \quad \frac{q_{t+1}\alpha^H(1-\theta)}{1-\frac{\theta q_{t+1}\alpha^H}{1+r_t}} > \pi R_t^x,$$

then H-types do not buy risky financial assets.

So long as $q_{t+1}\alpha^H > 1 + r_t$, we know

(56) $\quad \dfrac{q_{t+1}\alpha^H(1-\theta)}{1-\dfrac{\theta q_{t+1}\alpha^H}{1+r_t}} > q_{t+1}\alpha^H.$

Hence, we prove $q_{t+1}\alpha^H > \pi R_t^x$.

When we consider $D \to 0$ and use (43), $q_{t+1}\alpha^H > \pi R_t^x$ will become

(57) $\quad q_{t+1}\alpha^H > \pi \dfrac{P_{t+1}}{P_t} = \pi \dfrac{\theta q_{t+1}\alpha^H(1-\phi)}{\pi(1-p)-\phi}.$

By substituting (53) into (57), (57) becomes

(58) $\quad 1 - \pi\beta(1-p) > \pi\beta[1 - \pi(1-p)] + \pi\theta(1-\beta).$

Since the right-hand side is an increasing function of $\theta$ and the upper bound is $\theta = \pi\beta(1-p)$, substituting it into (58) yields

(59) $\quad (1 - \pi\beta)[1 - \pi\beta(1-p)] > 0,$

which is true. Hence, for any value of $\theta \in [\dfrac{\alpha^L}{\alpha^H}(1-p), \pi\beta(1-p))$, H-types do not buy risky financial assets.